\documentclass[letterpaper,twocolumn,10pt]{article}
\usepackage{usenix_2020_09}
\usepackage{hyperref}
\usepackage{xurl}

\usepackage{tikz}
\usepackage{gnuplottex}
\usepackage{epstopdf}
\usepackage{amsthm, amsmath,amssymb, wasysym}
\usepackage{bm}

\usepackage{appendix}

\usepackage{filecontents}

\usepackage{comment}
\usepackage{color}

\usepackage{booktabs}
\usepackage{multirow}
\usepackage{graphicx}

\usepackage{subcaption}

\theoremstyle{definition}
\newtheorem{definition}{Definition}[]

\usepackage{siunitx}

\usepackage{enumitem}
\setitemize[1]{itemsep=5pt, partopsep=0pt, parsep=\parskip,topsep=5pt}

\begin{document}

\date{}

\title{\Large \bf WFDefProxy: Modularly Implementing and Empirically Evaluating\\ Website Fingerprinting Defenses}

\author{
{\rm Jiajun Gong\textsuperscript{1}, Wuqi Zhang\textsuperscript{1}, Charles Zhang\textsuperscript{1}, Tao Wang\textsuperscript{2}}\\
\textsuperscript{1}The Hong Kong University of Science and Technology\\
\textsuperscript{2}Simon Fraser University\\
{\rm \{jgongac, wzhangcb, charlesz\}@cse.ust.hk, taowang@sfu.ca}
} 

\maketitle

\begin{abstract}
Tor, an onion-routing anonymity network, has been shown to be vulnerable to Website Fingerprinting (WF), 
which de-anonymizes web browsing by analyzing the unique characteristics of the encrypted network traffic. 
Although many defenses have been proposed, few have been implemented and tested in the real world;
others were only simulated.
Due to its synthetic nature, simulation may fail to capture the real performance of these defenses.
To figure out how these defenses perform in the real world, we propose WFDefProxy, a general platform for WF defense implementation on Tor using pluggable transports.
We create the first full implementation of three WF defenses: FRONT, Tamaraw and Random-WT.
We evaluate each defense in both simulation and implementation to compare their results, and we find that
simulation correctly captures the strength of each defense against attacks. 
In addition, we confirm that Random-WT is not effective in both simulation and implementation,
reducing the strongest attacker's accuracy by only 7\%.

We also found a minor difference in overhead between simulation and implementation. 
We analyze how this may be due to assumptions made in simulation regarding packet delays and queuing,
or the soft stop condition we implemented in WFDefProxy to detect the end of a page load. 
The implementation of FRONT cost about 23\% more data overhead than simulation,
while the implementation of Tamaraw cost about 28\% -- 45\% less data overhead.
In addition, the implementation of Tamaraw incurred only 21\% time overhead, compared to 51\% -- 242\% estimated by simulation in previous work.
\end{abstract}

\section{Introduction}

As people's awareness of the threat of tracking and surveillance grows, Tor~\cite{DingledineMS04}, the onion routing anonymity network, has become a popular tool to protect user privacy on the Internet. 
However, \textit{website fingerprinting} (WF), a kind of traffic analysis attack, has been shown to be able to de-anonymize Tor users.
WF attacks make use of timing, ordering and size features of network traffic traces to identify the webpages being visited, without breaking the encryption.
Recently, deep learning based attacks have achieved over 98\% accuracy~\cite{SirinamIJW18, bhat2019var} on a large multi-class de-anonymization task. 
High accuracy and unobservability make WF a serious threat to user privacy. 

Since the first WF attack was proposed, quite a few defenses have been put forward. 
Most prior defenses, especially those that function on the network layer, were tested and presented using simulation:
defense-less data is collected on the real Tor network, and then modified based on the theoretical effect of the defense.
Simulation allows the experimenter to measure defense performance cheaply, and without burdening the real Tor network.
However, simulation fails to evaluate some issues like the difficulty of implementation and
how the defense should cope with negative network events such as congestion. 
These issues may lead to inaccurate simulation results.
To this date, only randomized pipelining~\cite{Ranpipe} has been fully implemented on Tor, but it was shown to be ineffective and later disabled~\cite{Cai12touching}.
Many other defenses that claim to be effective and implementable have no prototypes in the real Tor network that can be tested and used. 
It remains to be seen how these defenses perform in reality.

It is necessary to figure out how these defenses perform on the real network before they can be recommended for adoption in Tor.
In this work, we build a platform called \textit{WFDefProxy} based on obfs4proxy~\cite{obfs4}, a widely-used pluggable transport that helps circumvent censorship.
On the platform, we implement two state-of-the-art defenses, FRONT~\cite{GongW20} and Tamaraw~\cite{CaiNWJG14}, as well as an unevaluated defense Random-WT~\cite{WangG17}.
In our implementation, we address a number of practical issues that were ignored in simulation such as how to decide the stop and end of a page load.
We directly collect several defended datasets using this framework and test each defense against the WF attacks. 

We summarize our contributions as follows:
\begin{itemize}
 \item We create a general platform called WFDefProxy for WF defense implementation on Tor using pluggable transports. 
 WFDefProxy is powerful enough to implement any known network-layer WF defense and should prove useful for WF research. 
 We create the first full implementation of three WF defenses on the platform, allowing real Tor users to adopt these defenses.
 \item We collect several datasets using WFDefProxy (each has up to 100,000 instances) and evaluate these three WF defenses under both simulation and implementation. Our results show that simulation results can be different from implementation ones, especially in the estimation of defense overhead, and we seek to explain why. 
 \item We make the first full evaluation for Random-WT, which has never been tested before, and show that it is not effective against deep learning based attacks.
\end{itemize}

We organize the paper as follows. 
We provide the background to our work in Section~\ref{sec:background} and introduce related work in Section~\ref{sec:related-work}. 
We present our new WF defense platform WFDefProxy and introduce the details of the implementations for three defenses in Section~\ref{sec:implementing-wf-defenses}.
We evaluate the implemented defenses extensively in Section~\ref{sec:implementation-evaluation}.
We compare the simulation and evaluation results in Section~\ref{sec:simulation-evaluation}.
Finally, we discuss relevant issues in Section~\ref{sec:discussion} and conclude our work in Section~\ref{sec:conclusion}.

\section{Background}
\label{sec:background}
In this section, we introduce some background knowledge about website fingerprinting and define our threat model. 

\subsection{Attack Scenario}
Website Fingerprinting (WF) can be viewed as a multi-class problem of classifying packet traces to webpages. 
The attacker \textit{monitors} a number of webpages of interest,
and tries to identify which monitored webpage each packet trace belongs to.
We usually consider two different attack scenarios in WF: the closed-world scenario and the open-world scenario.

The \textit{closed-world} scenario assumes that the client only visits webpages from a set of monitored webpages~\cite{HerrmannWF09}.
The attacker is trying to infer which specific page the client is visiting. 
This is the worst-case scenario for the client (and thus the defense), 
as it assumes the attacker has full knowledge of a client's possible browsing destinations.
We use \textit{accuracy}, the percentage of instances that are correctly classified, to evaluate the attacker's performance in this scenario.

The \textit{open-world} scenario is more realistic in that the client not only visits monitored webpages, 
but also visits non-monitored webpages --- webpages that the attacker is not interested in or has not seen before~\cite{PanchenkoNZE11}. 
The attacker's goal is to determine whether or not the client is visiting a webpage from his monitored list (binary-class problem), and if so,
to further answer which one (multi-class problem).
This is more challenging for the attacker compared to the closed-world scenario. 
We mainly use \textit{True Positive Rate (TPR)} and \textit{False Positive Rate (FPR)} to evaluate an attack in the open-world scenario.
A true positive is defined as a correctly classified instance that belongs to a monitored webpage. 
A false positive is defined as an instance that belongs to a non-monitored page and is misclassified as a monitored one. 
TPR is the percentage of true positives to the total number of monitored instances.
FPR is the percentage of false positives to the total number of non-monitored instances.

\subsection{Threat Model}
\label{sec:threat_model}
\begin{figure}
	\centering
	\includegraphics[width=\linewidth]{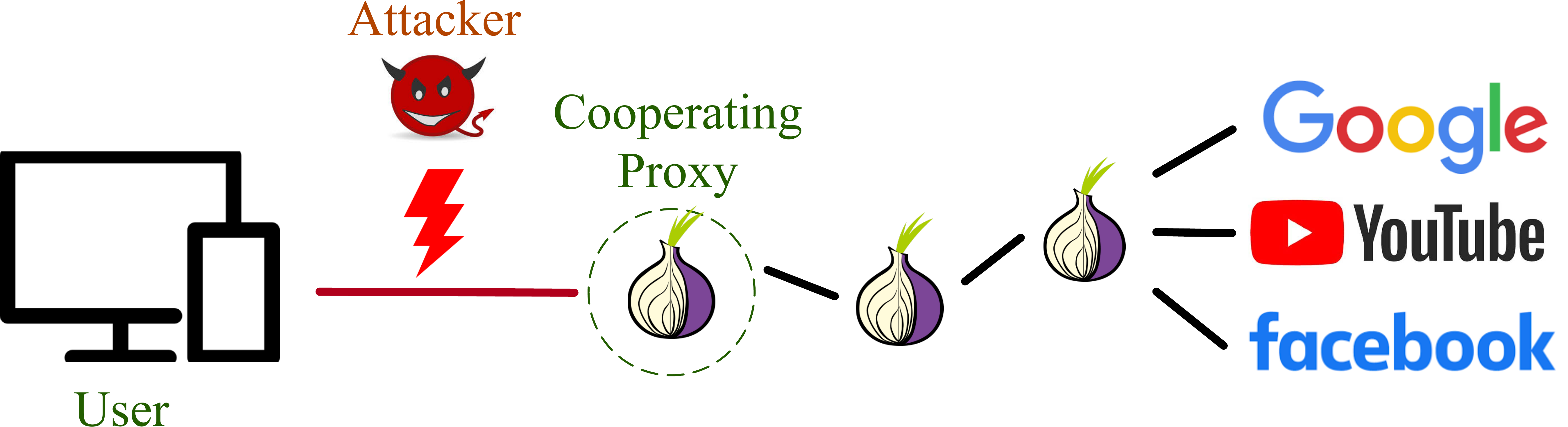}
	\caption{WF threat model.}
	\label{fig:attack-model}
\end{figure}
Figure~\ref{fig:attack-model} illustrates the threat model. 
There are three roles in the model: 

\begin{itemize}
\item \textit{User}: 
a user refers to someone who visits webpages through the Tor network. 
They may use a defense to protect themselves from potential WF attacks.
We use the terms \textit{user} and \textit{client} interchangeably. 

\item \textit{Attacker}: 
also referred to as the \textit{adversary}.
The adversary sits between the user and the entry node, passively eavesdropping on the connection.
The adversary does not attempt to compromise Tor's encryption;
instead, they try to infer the webpage being visited by analysing network traffic patterns with a WF classifier.
Anyone under the same network as the user or the Internet Service Provider could be a potential WF attacker.

\item \textit{Cooperating proxy}: 
The cooperating proxy is a Tor relay that helps deploy a network layer defense with the user. 
It will insert or drop dummy packets or delay real packets, depending on the defense protocol. 
Due to a technical limitation, we implement the defenses as Pluggable Transports (PT)~\cite{PT} on
the first node of the Tor circuit rather than the middle node.
\end{itemize} 

We further assume that the attacker is aware of the user's defense,
including the parameter settings of the defense.
This is a reasonable assumption for a local attacker since the attacker can easily derive them by observing client traffic.

\subsection{Trace and Overhead}
\label{sec:overhead}

A \textit{trace} is a sequence of network packets generated by loading a webpage. 
Data overhead and time overhead are the two main metrics we use to evaluate the cost of a defense.
\begin{definition} [Data Overhead]
	Let $w \in W$ be a trace and $\mathcal{D}(w)$ be the corresponding trace under defense $\mathcal{D}$, where $W$ is the set of all traces. 
	The data overhead of $\mathcal{D}$ is 
	$$DO(\mathcal{D}) = \mathbb{E}_{w\in W}\left(\frac{|\mathcal{D}(w)| - |w|}{|w|}\right),$$
	where $\mathbb{E}$ denotes the mean of its variable in the given set, and $|\cdot|$ denotes the total bytes of a trace. 
	
	Data overhead measures the amount of extra data required, which could affect throughput and burden the anonymity network.
\end{definition}

\begin{definition} [Time Overhead]
Let $t(\cdot)$ be the function that outputs the loading time of a trace, i.e., the timestamp of the last \textbf{real} packet of trace $w$.~\footnote{We specify that only the last real packet contributes to time overhead as some defenses add dummy packets after the end of real traffic.}
The time overhead of $\mathcal{D}$ is 
$$
TO(\mathcal{D}) = \mathbb{E}_{w\in W}  \left(    \frac{t(\mathcal{D}(w))-t(w)}{t(w)}   \right).
$$

Time overhead shows how much longer it will take for the user to load a web page;
it has a significant effect on the user's browsing experience. 
\end{definition}

An interesting point to note is that data overhead and time overhead are defined independently from each other.
In previous work that uses simulation, it is assumed that extra data overhead causes no extra time overhead.
In other words, the connection between the client and the cooperating proxy is assumed to have infinite bandwidth.
Since we will obtain results directly from implementation, we will be able to avoid this assumption.

\section{Related Work}
\label{sec:related-work}
\begin{table*}[]
\centering
\caption{Summary of existing defenses. We show the defenses we are going to implement in bold font: Tamaraw and FRONT are the state-of-the-art defenses in their category. Random-WT is a variant of Walkie-Talkie that is never evaluated before.}
\vspace{2pt}
\label{tab:defense-survey}
\resizebox{\textwidth}{!}{%
\begin{tabular}{|l||c|c|c|c|c|}
\hline
\multicolumn{1}{|c||}{\textbf{Category}} &
  \textbf{Defense} &
  \textbf{Function Layer} &
  \textbf{Evaluation Method} &
  \textbf{Requirement} &
  \textbf{Defeated by} \\ \hline \hline
\multirow{7}{*}{Non-regularized defenses} &
  Randomized Pipelining (2011)~\cite{Ranpipe} &
  Application &
  Implementation &
  None &
  DLSVM~\cite{Cai12touching} \\
 &
  HTTPOS (2011)~\cite{LuoZCLCP11} &
  Application &
  Implementation &
  None &
  kNN~\cite{WangCNJG14} \\
 &
  ALPaCA (2017)~\cite{CherubinHJ17} &
  Application &
  Implementation &
  Server-side cooperation &
  None \\
 &
  LLaMA (2017)~\cite{CherubinHJ17} &
  Application &
  Implementation &
  None &
  None \\
 &
  WTF-PAD (2016)~\cite{JuarezIPDW16} &
  Network &
  Simulation &
  None &
  DF~\cite{SirinamIJW18} \\
 &
  \textbf{FRONT (2020)~\cite{GongW20}} &
  \textbf{Network} &
  \textbf{Simulation} &
  \textbf{None} &
  \textbf{None} \\
 &
  TrafficSliver (2020)~\cite{Cadena20Traffic} &
  Network &
  Implementation &
  Multiple Entry Nodes &
  None \\ \hline
\multirow{7}{*}{Regularized defenses} &
  BuFLO (2012)~\cite{DyerCRS12} &
  Network &
  Simulation &
  None &
  None \\
 &
  CS-BuFLO (2014)~\cite{CaiNJ14} &
  Network &
  Implementation &
  None &
  None \\
 &
  \textbf{Tamaraw (2014)~\cite{CaiNWJG14}} &
  \textbf{Network} &
  \textbf{Simulation} &
  \textbf{None} &
  \textbf{None} \\
 &
  Glove (2014)~\cite{NithyanandCJ14} &
  Network &
  Simulation &
  Knowledge of pages &
  None \\
 &
  Supersequence (2014)~\cite{WangCNJG14} &
  Network &
  Simulation &
  Knowledge of pages &
  None \\
 &
  Walkie-Talkie (2017)~\cite{WangG17} &
  Network &
  Simulation &
  Knowledge of pages, half-duplex mode &
  None \\
 &
  \textbf{Random-WT (2017)~\cite{WangG17}} &
  \textbf{Network} &
  \textbf{N/A} &
  \textbf{Half-duplex mode} &
  \textbf{N/A} \\ \hline
\end{tabular}%
}
\end{table*}

In this section, we survey the existing works on website fingerprinting attacks and defenses.
We further systematically categorize WF defenses and we show the necessity to implement the state-of-the-art defenses.

\subsection{Website Fingerprinting Attacks}

A WF attacker can guess which webpage the client is visiting 
because different webpages generate different traffic patterns.
Hintz~\cite{Hintz02} exploited the size of packets received to fingerprint webpages.
Sun et al.~\cite{SunSWRPQ02} identified webpages by examining the bytes of chunks received since they reveal the size and number of objects in a webpage. 
Later, machine learning algorithms were introduced to facilitate website fingerprinting. 
The Na\"ive-Bayes Classifier was shown to be effective in attacking single hop systems like OpenSSH and VPN~\cite{HerrmannWF09, LiberatoreL06}. 
Panchenko et al.~\cite{PanchenkoNZE11} presented the first successful attack on Tor using a SVM classifier. 
After that, more attacks were proposed, using different machine learning models with more sophisticated feature engineering processes~\cite{WangG13improve, Cai12touching, Panchenko16Web, Hayes16kfin, WangCNJG14}, greatly improving the performance of WF attacks on Tor. 
To automate the feature extraction process, several deep learning methods were proposed~\cite{SirinamIJW18, abe2016fingerprinting, bhat2019var, RimmerPJGJ18Automated}.
One such attack, Deep Fingerprinting~\cite{SirinamIJW18}, 
even achieved over 90\% accuracy in a closed-world scenario of 95 websites against WTF-PAD~\cite{JuarezIPDW16}, 
which had been a candidate defense for deployment on Tor~\cite{wtfpad,wtfpad-2}. 
Currently, the deep learning based classifiers are the most powerful WF attacks.

\subsection{Website Fingerprinting Defenses}
Over the years, researchers have proposed a number of WF defenses in response to increasingly powerful WF attacks. 
We can generally divide WF defenses into two main categories: 
regularized defenses and non-regularized defenses.

Regularized defenses predefine a set pattern for traffic traces to be molded into.
They usually focus on achieving strong security, 
so they tend to incur a high overhead and cause significant delays to user traffic.
For example, BuFLO-family defenses~\cite{DyerCRS12, CaiNWJG14, CaiNJ14} limit the packet sending rate and pad the length of the traffic.
Under these defenses, webpages of a similar size will often generate the exact same network traffic, leaving little information to the attacker.
Among BuFLO-family defenses, only CS-BuFLO~\cite{CaiNJ14} has been fully implemented and tested,
and it was shown to cost more than 200\% data and time overhead.
Tamaraw~\cite{CaiNWJG14} was further optimized on CS-BuFLO by setting different packet sending rates for outgoing and incoming packets and improving the padding scheme at the end of each page load.~\footnote{We define packets sent from the client side as \textbf{outgoing} packets.}
It claimed to be more efficient than CS-BuFLO through simulation.
However, different studies have reported significantly different overhead values for Tamaraw~\cite{WangG17, SirinamIJW18, bhat2019var, GongW20}.
The minimum overhead reported among these works is 63\% in data and 51\% in time while the maximum is 328\% in data and 242\% in time.

There is another class of regularized defenses in which webpages are grouped into different anonymity sets so that traces generated by webpages in one set appear the same~\cite{WangCNJG14, NithyanandCJ14, WangG17}. 
These defenses all require some prior knowledge of webpages to facilitate anonymization.
Walkie-Talkie~\cite{WangG17} requires the least overhead among these defenses. 
Wang and Goldberg implemented a customized Tor Browser which is able to run in half-duplex mode and simulated Walkie-Talkie on it.
However, Walkie-Talkie requires a database that stores the burst sequences of a large number of webpages to do burst molding~\cite{WangG17}, 
and this has proven hard to achieve in reality~\cite{mohammadTik19}.
The authors also put forward a relaxed version of Walkie-Talkie that does not require burst molding, Random-WT, that pads real bursts and randomly adds fake bursts.
Random-WT has not been evaluated against the latest attacks. 

Non-regularized defenses do not guarantee their effectiveness against all attacks, but require much less overhead. 
They often obfuscate network traces by injecting dummy packets to alter certain characteristics of the traffic, trying not to delay real user packets.
For example, 
Randomized Pipelining~\cite{Ranpipe} and HTTPOS~\cite{LuoZCLCP11} modify (reorder, delay or resize) HTTP requests.
Cherubin et al.~\cite{greschbach2017effect} proposed two application layer defenses, a server-side defense ALPaCA and a client-side defense LLaMA.
They change the shape of the loading process by padding the request or response of each resource with dummy bytes.
WTF-PAD~\cite{JuarezIPDW16} and FRONT~\cite{GongW20} are two zero-delay defenses that randomly insert dummy packets into the traces. 
Recently, Cadena et al.~\cite{Cadena20Traffic} proposed a new defense called TrafficSliver which split traffic over multiple entry nodes. 
It is shown to effectively counter any single malicious entry node with little overhead. 
However, it cannot defend against any local attacker who is able to see the whole traffic before splitting.

We summarize the defenses discussed in Table~\ref{tab:defense-survey}.
Most defenses were evaluated only under simulation and not implementation. 
Out of those, we pick three that we believe have the most potential to succeed:
FRONT, the newest non-regularized defense; Tamaraw, a strong regularized defense;
and Random-WT, a practical substitute for Walkie-Talkie.
We implement these defenses and evaluate them with both simulation and implementation.

\section{Implementing WF Defenses}
\label{sec:implementing-wf-defenses}
In this section, we introduce WFDefProxy, a platform we created for implementing WF defenses. 
We show, in detail, how we design each defense as a pluggable transport on this platform 
and describe how the implementations differ from previous simulations.

\subsection{The WFDefProxy Platform}
\label{sec:the-wfdefproxy-platform}
As shown in Table~\ref{tab:defense-survey}, most defenses functioning at the network layer remain unevaluated. 
Juarez once developed a WFPadTools framework to facilitate WF defense research in 2015~\cite{wfpadtools},
but it does not provide encryption and is no longer maintained. 
Recently, the Tor Project developed a Circuit Padding Framework that can implement circuit-level padding defenses in Tor~\cite{cirpad}.
However, it does not support delaying packets due to the risk of causing out-of-memory problems on Tor relays.
Therefore, it is impossible to directly implement many defenses including Random-WT and Tamaraw in this framework.
Although FRONT does not delay any packets, it is still hard to directly implement it in the framework because the framework utilizes a state machine that acts on ``packet-level'' events instead of ``trace-level'' ones, which are required by FRONT (See Section~\ref{sec:implementing-front}). 
We develop a new platform, WFDefProxy, to implement these defenses. 

\paragraph{Introduction}
WFDefProxy is a pluggable transport proxy for Tor.
Pluggable transports were originally designed to circumvent censorship in countries that banned Tor:
the user cooperates with the bridge to transform the Tor traffic into innocuous-looking traffic. 
WFDefProxy extends obfs4proxy, a pluggable transport that implements the latest well-evaluated obfuscation protocol obfs4~\cite{obfs4}.
WFDefProxy utilizes the cryptography module of obfs4proxy.

\begin{figure}
	\centering
	\includegraphics[width=0.9\linewidth]{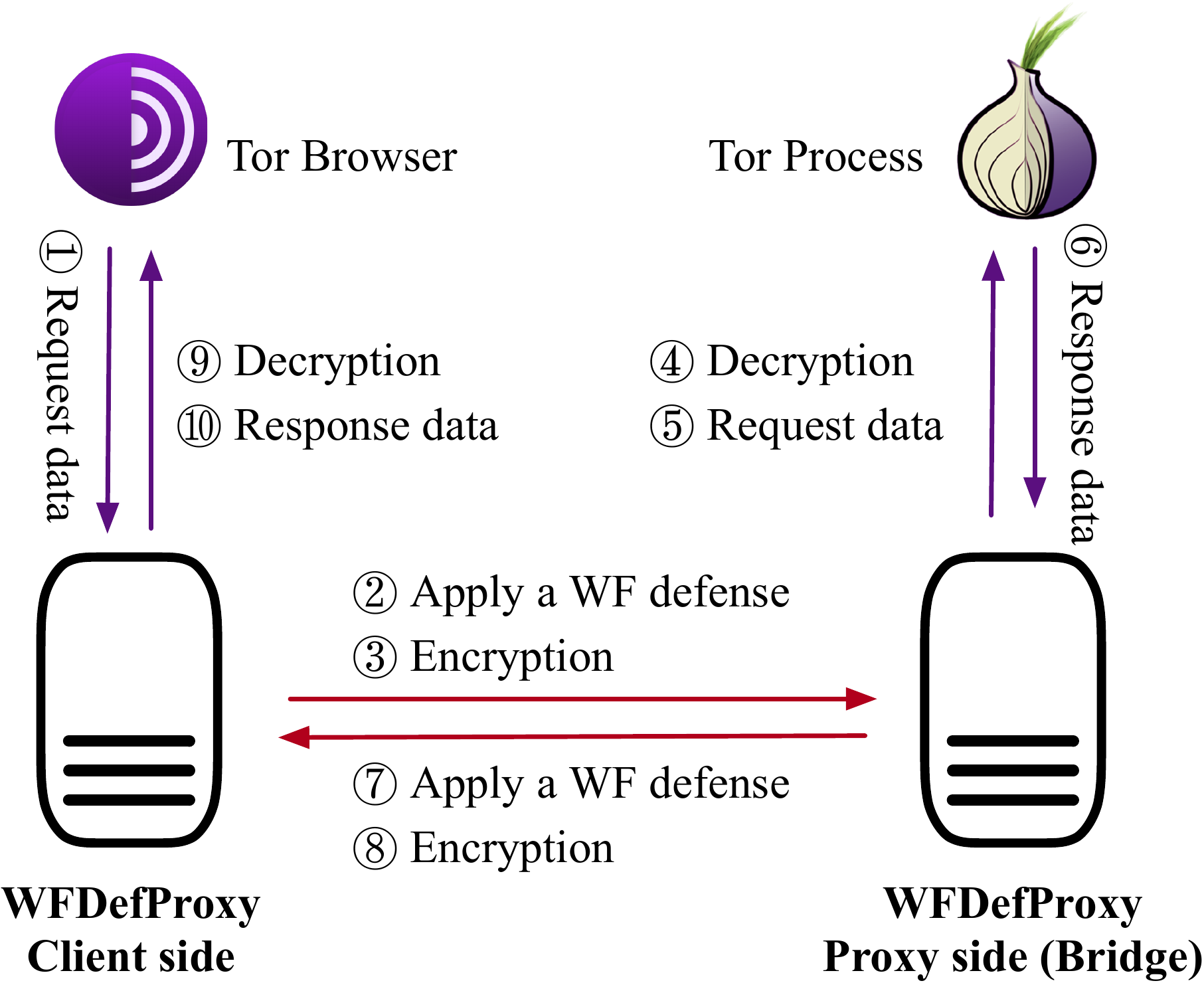}
	\caption{The workflow of WFDefProxy. WFDefProxy adds another layer of encryption (denoted in red arrows) upon Tor's encryption (denoted in purple arrows) so that the attacker is unable to distinguish between real and dummy packets.}
	\label{fig:workflow-of-wfdefproxy}
\end{figure}

Figure~\ref{fig:workflow-of-wfdefproxy} illustrates the workflow of WFDefProxy. 
Data from Tor Browser (the Tor process at the proxy side) will first be sent to the WFDefProxy platform.
According to the defense used, the WFDefProxy will modify the data, delaying real packets and/or adding dummy packets.
After that, another layer of encryption will be added to all packets before they are sent onto the wire,
so that the attacker will be unable to distinguish between real and dummy packets. 
After these packets arrive on the other side, they will be decrypted and forwarded to the Tor layer. 

We have implemented three defenses on the platform so far: FRONT~\cite{GongW20}, Tamaraw~\cite{CaiNWJG14} and Random-WT~\cite{WangG17}.
To run one of the defenses, the bridge should add the option \texttt{ServerTransportPlugin} in the \texttt{torrc} file~\cite{torrc}.
After launch, it will generate a file containing a certificate needed for the handshake as well as the corresponding parameters used for the defense.
The client should provide the certificate and the correct parameters when connecting to the bridge. 
After a successful handshake, the traffic will be obfuscated based on the requested defense protocol.

\paragraph{Practical Issues}
A major difference between simulation and implementation for a WF defense is that in simulation,
the defense knows ahead of time when it will finish loading a webpage.
However, in implementation, it is non-trivial for a network-layer defense to know when a page has finished loading. 
A conservative stop strategy may make the defense send more dummy data than necessary, which wastes bandwidth.
On the other hand, aggressively stopping page loading would cause early stops, which may leak information and thus weaken the defense.

To tackle this problem, we estimate the end of loading by observing the throughput of real packets over a short period of time.
When the throughput is close to 0 in the time window, we assume page loading has finished. 
We refer to this as the \textbf{soft stop condition}. 
The intuition is that when a page is fully loaded, we should observe few real packets on the connection.
Even though it is still possible to see some packets (such as those related to Tor's link padding protocol and flow control protocol~\cite{torspec}), those packets would not occur frequently. 
To decide the window size, we measure the time gaps between two consecutive outgoing packets on the undefended dataset we collected in Section~\ref{sec:implementation-evaluation}. 
We find that 99\% of time gaps are less than $\SI{1}{s}$.
Therefore, we set a time window of $\SI{1}{s}$ to decide when a page load ends.
If there is no more than one packet over the last $\SI{1}{s}$, we stop the defense.

Another difference is that simulation does not consider necessary information exchange between the client and the proxy. 
The proxy needs to synchronize with the client from time to time so that they are in the same state of the defense.
To tackle this problem, we define a special signal packet to facilitate communication between the two parties so that both sides can start and stop padding together as expected.

\subsection{Implementing Tamaraw}
\label{sec:implementing-tamaraw}
\begin{table}[]
\centering
\caption{The parameters of Tamaraw and their default values.}
\label{tab:tamaraw-params}
\resizebox{\linewidth}{!}{%
\begin{tabular}{|c||c|c|}
\hline
\textbf{Parameter} & \textbf{Value} & \textbf{Description}                           \\ \hline\hline
$\rho_{out}$       & $\SI{12}{ms}$ & Packet sending gap at the client side          \\ \hline
$\rho_{in}$        & $\SI{4}{ms}$  & Packet sending gap at the server side          \\ \hline
$L$                & 200           & Trace length is padded to the multiples of $L$ \\ \hline
\end{tabular}%
}
\end{table}

Tamaraw is a regularized defense that can provide a theoretical upper bound on the accuracy of any attacker. 
It is widely used as a baseline comparison with other defenses and to test WF attacks.
Its parameters are listed in Table~\ref{tab:tamaraw-params}.
Packets are sent every $\rho_{out}$ milliseconds on the client side and $\rho_{in}$ milliseconds on the proxy side. 
Dummy packets are inserted if no data can be sent.
When loading finishes, Tamaraw continues to pad the flow until the length of the trace is a multiple of $L$. 
We set $\rho_{out} = \SI{12}{ms}, \rho_{in} = \SI{4}{ms}$ and $L = 200$ in our implementation. 
We slightly decrease $\rho_{out}$ and $\rho_{in}$ compared to the values in the original work~\cite{CaiNWJG14} since their payload is defined as 750 bytes which is slightly larger than ours (514 bytes).

\begin{figure*}
    \centering
    \begin{subfigure}{.325\textwidth}
      \centering
      \includegraphics[width=\linewidth]{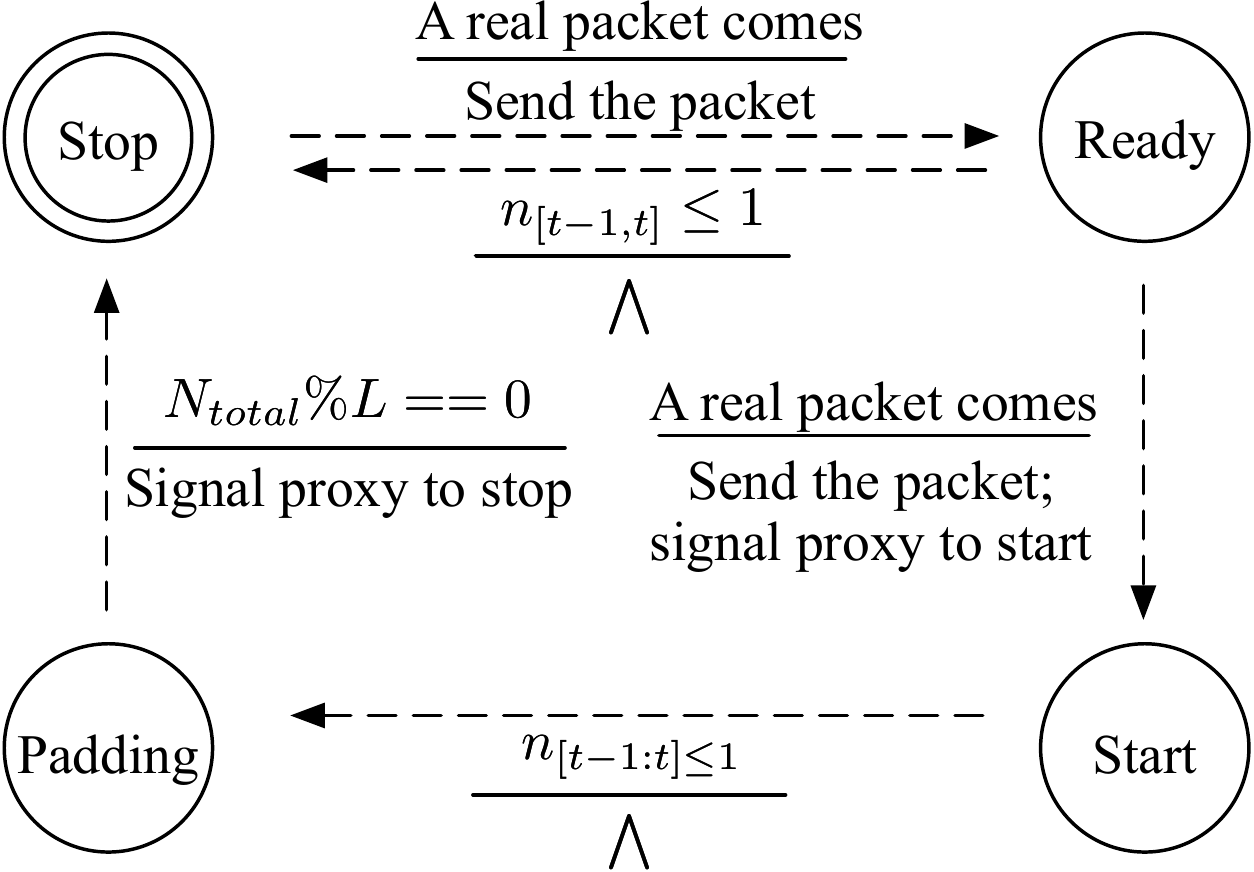}  
      \caption{Tamaraw}
      \label{fig:tamaraw-fsm}
    \end{subfigure}
    \hspace{1pt}
    \begin{subfigure}{.325\textwidth}
      \centering
      \includegraphics[width=\linewidth]{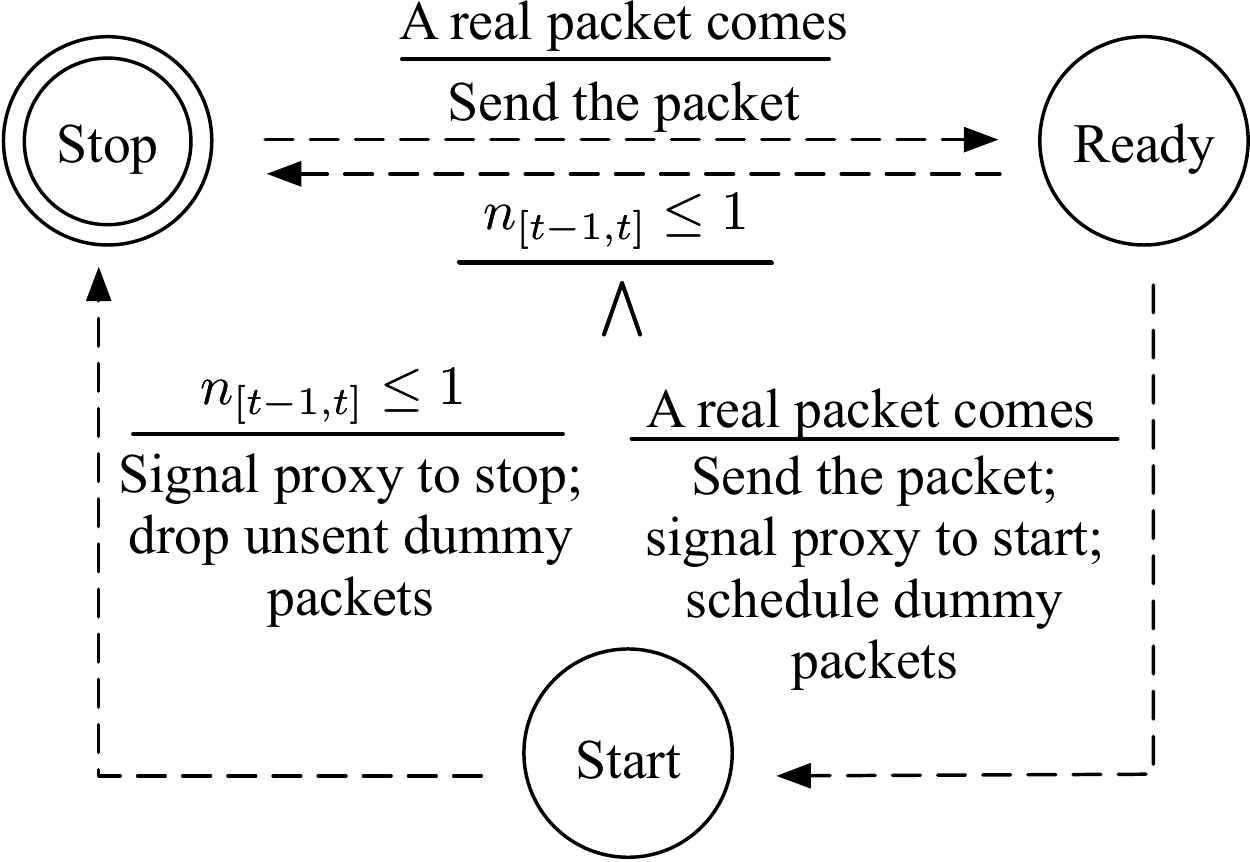}  
      \caption{FRONT}
      \label{fig:front-fsm}
    \end{subfigure}
     \hspace{1pt}
    \begin{subfigure}{.325\textwidth}
      \centering
      \includegraphics[width=\linewidth]{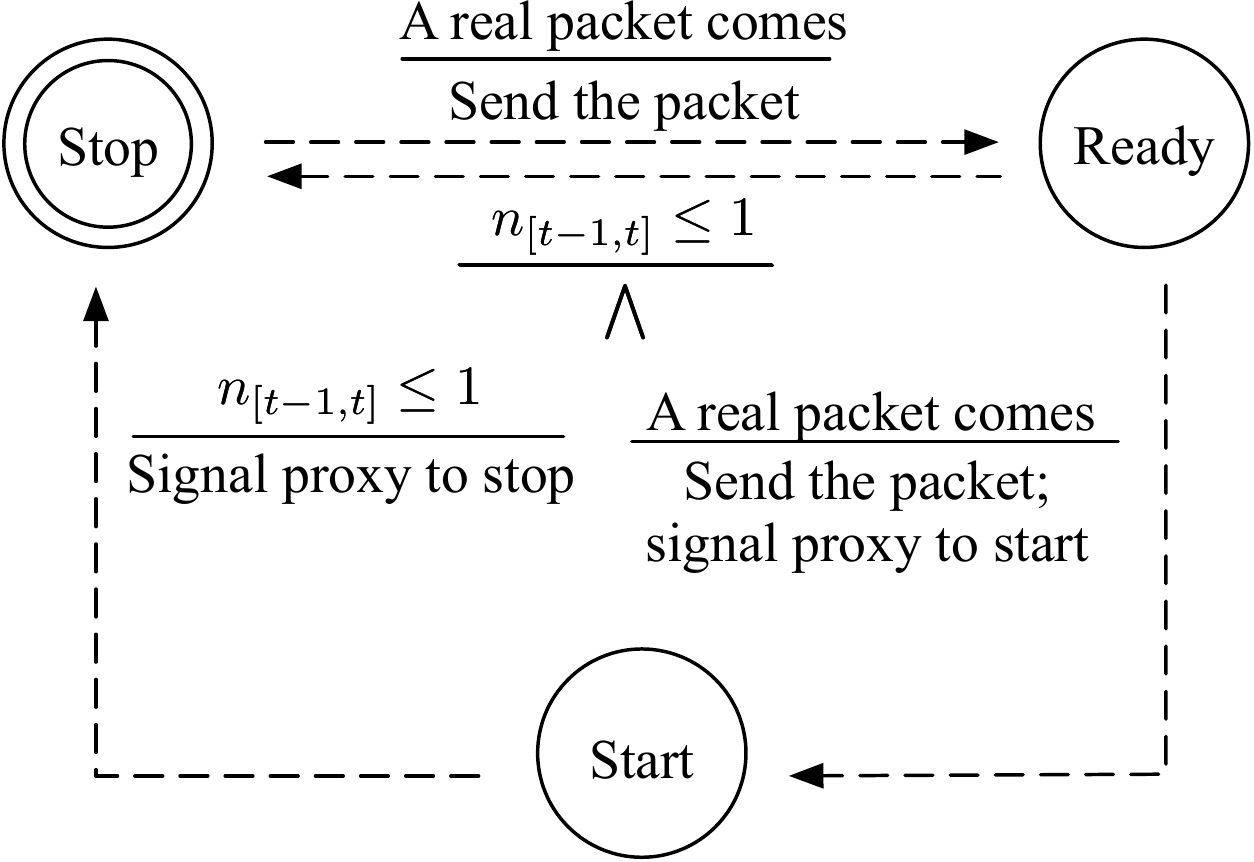}  
      \caption{Random-WT}
      \label{fig:randomwt-fsm}
    \end{subfigure}
    \caption{The state machine on the client side for each defense. 
    Above each line, there is an event that triggers the actions below the line and causes a state transition. 
    For simplicity, we omit events that do not cause a state transition.
    $``\bigwedge"$ means no action. $n_{[t-1:t]}$ refers to the number of real packets sent over the past second.
    The padding for each defense happens in the Start and Padding states.}
    \label{fig:fsm}
\end{figure*}

We use a finite state machine to control the behavior of Tamaraw.
Figure~\ref{fig:tamaraw-fsm} depicts the state transition on the client side. 
Each circle denotes a state and the double circle denotes the initial state when the defense starts. 
We use $n_{[t-1:t]} (\forall t \geq 1$) to denote the number of \textbf{real} packets over the past second. 
As discussed in Section~\ref{sec:the-wfdefproxy-platform}, we use this number to detect the end of a page load.
$N_{total}$ denotes the total number of packets, including the dummy ones.
For simplicity, we only show the events that will cause a transition between two different states. 
For all other cases, the defense will send real or dummy packets at suitable time slots according to the protocol of Tamaraw, staying in the same state. 

There are four states in the client's state machine: Stop, Ready, Start, and Padding. 
The defense will inject dummy packets in the Start and Padding states. 
We introduce a Ready state before the Start state as a buffer:
if there is only one packet over a time window of $\SI{1}{s}$ when the machine is in the Ready state, it will return to the Stop state.
This is used to avoid frequent triggering of padding by non-payload packets since Tamaraw padding is quite expensive. 
The proxy's state machine does not change state autonomously;
rather, it will be signaled to start Tamaraw padding when the client enters the Start state,
and to stop padding when the client enters the Stop state.


\subsection{Implementing FRONT}
\label{sec:implementing-front}
FRONT is a lightweight defense that does not delay any real packets.
The idea behind FRONT is to add more randomness into the defense and to put more effort into obfuscating the trace front --- the first few seconds of page loading.
It achieves this by randomizing the number of dummy packets injected into the trace, and the timestamps of these dummy packets are sampled from a Rayleigh distribution. 

As shown in Table~\ref{tab:front-params}, there are four parameters. 
When a loading begins, each side will first sample an integer $n$ from $\bar{U}(1, N_{r})$, $r \in \{out, in\}$, where $\bar{U}(a, b)$ denotes a discrete uniform distribution from $a$ to $b$.
Then $n$ timestamps will be sampled from a Rayleigh Distribution $\mathcal{R}(t;w) = \frac{t}{w^2} e ^ {-t^2 / 2 w^2}$, 
where $w$ is sampled from $U(W_{min}, W_{max})$, where $U(a, b)$ is the continuous uniform distribution from $a$ to $b$.
$w$ is a parameter that controls the shape of the distribution.
There are expected to be 40\% of the sampled timestamps that lie in the time interval [0, $w$].
This initialization process is called ``packet scheduling''.
These dummy packets will be sent at the sampled timestamps. 
After loading finishes, unsent dummy packets will be dropped. 
We set $N_{out} = N_{in} = 3000, W_{min} = \SI{1}{s}$ and $W_{max} = \SI{13}{s}$, close to the suggested settings in the original work~\cite{GongW20}.

Figure~\ref{fig:front-fsm} shows that there are three states in our implementation of FRONT: Stop, Ready, and Start.
Similar to Tamaraw's implementation, FRONT padding will not be easily triggered by noise unless there are at least two real packets over the past second. 
When the client enters the Start state, it will start padding according to the FRONT protocol and signal the proxy to start padding on its side.
When there are no more than two packets in the previous second, the client's machine will enter the Stop state to stop padding,
and similarly signal the proxy to stop the defense and drop any unsent dummy packets.

\begin{table}[]
\centering
\caption{The parameters of FRONT and their default values.}
\label{tab:front-params}
\resizebox{\linewidth}{!}{%
\begin{tabular}{|c||c|c|}
\hline
\textbf{Parameter} & \textbf{Value} & \textbf{Description}                     \\ \hline\hline
$N_{out}$          & 3000           & Max padding number at the client side     \\ \hline
$N_{in}$           & 3000           & Max padding number at the proxy side   \\ \hline
$W_{min}$          & $\SI{1}{s}$      & Rayleigh Distribution parameter \\ \hline
$W_{max}$          & $\SI{13}{s}$     & Rayleigh Distribution parameter \\ \hline
\end{tabular}%
}
\end{table}


\subsection{Implementing Random-WT}
\label{sec:implementing-random-wt}
Walkie-Talkie (WT) forces the client and server to talk in half-duplex mode,
so that a trace becomes a sequence of outgoing and incoming packet bursts --- a ``burst sequence''.
WT requires the client to know the shape of the burst sequence of each webpage in advance. 
Then, it pads the trace by merging it with another randomly chosen burst sequence so that loading these two corresponding webpages will generate exactly the same trace. 
This process is called ``burst molding''. 

The biggest challenge of deploying WT in practice is that 
the client needs to know the burst sequences of many webpages in advance.
However, webpages do not have fixed burst sequences, due to their dynamic nature, browser configurations, etc.~\cite{mohammadTik19}. 
Furthermore, it is difficult to maintain a burst sequence database and to keep it updated.
Recently, Rahman et al.~\cite{mohammadTik19} claimed to have the first full implementation of WT.
Unfortunately, they did not solve the problems mentioned above. 
The random nature of webpage loading may cause the PT to incorrectly end a burst from time to time, splitting the traffic into more bursts than needed.
Rahman et al.\ reported extremely high overhead values in both data and time, which may not reflect the real performance of WT. 

Given that deploying WT in the current Tor network may be unrealistic, we decided to implement a relaxed version of WT --- Random-WT.
Random-WT is practically deployable because it does not mold bursts and thus does not need to maintain a database of burst sequences.
As in WT, the two parties talk in half-duplex mode, taking turns to send bursts of data. 
A party that is about to send data (i.e.\ in ``Talkie'' mode) simply pads the real burst with $n^{real}_r$ dummy packets or 
sends a fake burst (a burst of dummy packets) of size $n^{fake}_r$ with probability $p_{fake}$.
$n_r^{real}$ is sampled from $\bar{U}(0, N_{r}^{real})$ and $n_r^{fake}$ is sampled from $\bar{U}(0, N_{r}^{fake})$, $r \in \{in, out\}$.
Since the original work had no recommended settings, we explore suitable parameter values in Section~\ref{sec:simulating-random-wt}.
The parameters of Random-WT and the default values we use are summarized in Table~\ref{tab:randomwt-params}.

\begin{table}[]
\centering
\caption{The parameters of Random-WT and their default values.}
\label{tab:randomwt-params}
\resizebox{\linewidth}{!}{%
\begin{tabular}{|c||c|c|}
\hline
\textbf{Parameter} & \textbf{Value} & \textbf{Description}                      \\ \hline \hline
$N_{out}^{real}$   & 4                  & Max padding number on client's real burst \\ \hline
$N_{in}^{real}$    & 45                 & Max padding number on proxy's real burst  \\ \hline
$N_{out}^{fake}$   & 8                  & Max padding number on client's fake burst \\ \hline
$N_{in}^{fake}$    & 90                 & Max padding number on proxy's fake burst  \\ \hline
$p_{fake}$         & 0.4                & The probability of inserting a fake burst \\ \hline
\end{tabular}%
}
\end{table}

The state machine of Random-WT is similar to that of FRONT, as shown in Figure~\ref{fig:randomwt-fsm}.
It has three states and padding only happens in the Start state.
Note that when the machine is in the Stop and Ready states, the two parties will still talk in half-duplex mode,
but there will be no dummy data in these two states. 
We set a timer $t_{talkie}$ of \SI{500}{ms} for any party in Talkie mode.
If there is no real data coming from upstream within $t_{talkie}$, that party will directly send out a signal packet and switch to ``Walkie'' mode to avoid deadlock.

\section{Implementation Evaluation}
\label{sec:implementation-evaluation}
In this section, we empirically evaluate WF defenses on the real Tor network. 
We collected an undefended dataset and a defended dataset for each defense. 
We compare these defenses against the state-of-the-art attacks in both closed-world and open-world scenarios. 

\subsection{Methodology}
\paragraph{Experiment Setup}
We deploy 3 servers on Microsoft Azure, 2 acting as clients and 1 as a Tor bridge. 
All clients have 4 CPU cores (2.3 GHz) and 16 GB memory, running on Ubuntu~18.04.4~LTS.
The bridge server has 1 CPU (2.3 GHz) and 2 GB memory, running on Debian~9.11. 
We place the client servers close to us and the bridge server far from us, globally; 
their exact locations are scrubbed for blind review.
The bridge will act as the cooperating WFDefProxy. 

We create 12 docker containers on the client servers to collect datasets in parallel. 
In each container, we use command lines to launch a Tor Browser (version 10.0.15) directly for each visit.
Once the loading is completed, we will wait for an extra $\SI{5}{s}$ on the page, 
after which the browser will be automatically closed with a Tampermonkey~\cite{tampermonkey} script.
We give at most $\SI{70}{s}$ for each page load, after which the browser will be closed immediately. 
Each visit uses a fresh new copy of the Tor Browser Bundle to remove the impact of the cache on the crawling process. 
The bridge runs a Tor process of version 0.4.4.5.  
Unless otherwise specified, we use the default settings presented in Section~\ref{sec:implementing-wf-defenses} for each defense for the rest of the paper.

\paragraph{Datasets}
The website list we choose for evaluation comes from the Tranco top 1 million list~\cite{LePochat2019} generated on 21st January, 2021.~\footnote{\url{https://tranco-list.eu}}
Tranco is a regularly updated research-oriented ranking that aggregates data from Alexa~\cite{alexa}, Umbrella~\cite{umbrella} and Majestic~\cite{majestic}.
It shares a large portion of the most popular domains with these three rankings and provides a more stable ranking in terms of website popularity.

We first remove the inaccessible URLs in the top 200 sites. 
We also only keep one version of localized websites such as \texttt{Google}. 
In the remaining sites, we choose the first 100 as the monitored webpages. 
We then use the following 90,000 sites starting from the 201st site in the list as the non-monitored webpages. 
We visit all the monitored webpages 120 times and all the non-monitored webpages once,
crawling the pages in a Round-Robin fashion as prior works do~\cite{SirinamIJW18, WangG13improve}. 
Then we remove the outliers for the monitored traces using the approach described in~\cite{Panchenko16Web}.
We keep 100 traces for each monitored webpage; in the end, each dataset contains 100,000 traces in total. 
We collect one dataset for each implemented defense following this methodology. 
We also collect two undefended datasets: one using the normal Tor Browser, 
and one under half-duplex mode, which is used to simulate Random-WT as a comparison with our implementation in Section~\ref{sec:simulation-evaluation}. 
The whole crawling process lasts for one month.  
 
\paragraph{WF Attacks}
We pick four state-of-the-art WF attacks, kFP~\cite{Hayes16kfin}, CUMUL~\cite{Panchenko16Web}, DF~\cite{SirinamIJW18}, and Var-CNN~\cite{bhat2019var} to evaluate the defenses. 
kFP uses Random Forests and k-Nearest Neighbour classifiers to perform the attack.
CUMUL makes use of a SVM classifier with a ``cumulative representation'' of the traces as input.
DF and Var-CNN are two deep learning based attacks that use convolutional neural networks.
They were shown to be more effective in previous work.
We use the parameters suggested in their papers. 
For the deep learning attacks, we set the input length at 10,000 and train each model for 30 epochs.
We perform 10-fold cross validation on each dataset and add up the results on each fold.

\paragraph{Ethical Considerations}
Throughout our experiments, our bridge is kept private so that no one can connect to our bridge. 
We write a script to automatically drive the browser and none of the visits are from real users. 
The dummy packets are transmitted only between the clients and our private bridge, so no dummy packets flow into the Tor network. 
We do not keep any real data generated by the page loads.

\subsection{Closed-World Scenario}
We first investigate the closed-world scenario, which assumes the client only visits webpages that the attacker is monitoring.

\paragraph{Overhead}
Table~\ref{tab:defense-imp-closed-ovhd} shows the time and data overhead of each defense. 
Tamaraw is the most expensive defense, costing 142\% data overhead and 30\% time overhead. 
However, the time overhead is much less than previously reported results (which were obtained with simulation)~\cite{CaiNWJG14, SirinamIJW18, GongW20}.
This suggests that Tamaraw will not impact user experience as much as previously expected.
Random-WT incurs a similar time overhead (37\%) as WT,
but it has a much higher data overhead (82\%). 
FRONT incurs the least overhead among the defenses, with 68\% data overhead and nearly 0\% time overhead. 

\begin{table}[]
\centering
\caption{The data and time overhead of each defense in the closed-world scenario (100 $\times$ 100 traces).}
\label{tab:defense-imp-closed-ovhd}
\resizebox{0.98\linewidth}{!}{%
\begin{tabular}{|c||c|c|}
\hline
\textbf{Defense $\mathcal{D}$} & \textbf{Data Overhead (\%)} & \textbf{Time Overhead(\%)} \\ \hline \hline
Tamaraw~\cite{CaiNWJG14} & 142 & 30 \\
FRONT~\cite{GongW20}     & 68  & 2  \\
Random-WT~\cite{WangG17} & 82  & 37 \\ \hline
\end{tabular}%
}
\end{table}

\paragraph{Attack Accuracy}

\begin{table}[]
\centering
\caption{Attack accuracy in the closed-world scenario against each defense (100 $\times$ 100 traces).}
\label{tab:defense-imp-closed-acc}
\resizebox{\linewidth}{!}{%
\begin{tabular}{|c||cccc|}
\hline
\multirow{2}{*}{\textbf{Defense $\mathcal{D}$}} & \multicolumn{4}{c|}{\textbf{Attack Accuracy (\%)}}                                                                \\
                                                & kFP~\cite{Hayes16kfin}     & CUMUL~\cite{Panchenko16Web} & DF~\cite{SirinamIJW18}     & Var-CNN~\cite{bhat2019var} \\ \hline\hline
Undefended                                      & \multicolumn{1}{c|}{89.46} & \multicolumn{1}{c|}{95.65}  & \multicolumn{1}{c|}{98.26} & 96.79                     \\
Tamaraw~\cite{CaiNWJG14}                        & \multicolumn{1}{c|}{7.14}  & \multicolumn{1}{c|}{8.80}    & \multicolumn{1}{c|}{17.17} & 7.42                      \\
FRONT~\cite{GongW20}                            & \multicolumn{1}{c|}{24.46} & \multicolumn{1}{c|}{17.45}  & \multicolumn{1}{c|}{62.63} & 59.79                     \\
Random-WT~\cite{WangG17} & \multicolumn{1}{c|}{62.88} & \multicolumn{1}{c|}{67.30} & \multicolumn{1}{c|}{91.55} & 80.54 \\ \hline
\end{tabular}%
}
\end{table}

Table~\ref{tab:defense-imp-closed-acc} shows the attack accuracy on original Tor and each implemented defense.
When there is no defense, kFP can achieve 89\% accuracy while the other attacks can achieve more than 95\% accuracy. 
Tamaraw is the most effective defense, lowering the accuracy of kFP, CUMUL and Var-CNN to less than 9\%.
DF still achieves about 17\% accuracy against Tamaraw. 
FRONT is quite effective against kFP and CUMUL, lowering their accuracy to less than 25\% and 18\% respectively,
but the deep learning based attacks still achieve more than 59\% accuracy. 
We find that Random-WT is rather ineffective against all the attacks. 
Although it has higher overhead values than FRONT, all attacks achieve over 62\% accuracy against Random-WT. 
The best attack, DF, even gets 91\% accuracy against Random-WT, which is only a 7\% drop in accuracy compared with the undefended case.
Our results show that Random-WT, which was designed as a possible replacement for Walkie-Talkie without the requirement of prior webpage knowledge,
is an insecure defense.
 
\paragraph{Information Leakage}
\begin{figure}[]
	\begin{center}
		\includegraphics[width=0.99\linewidth]{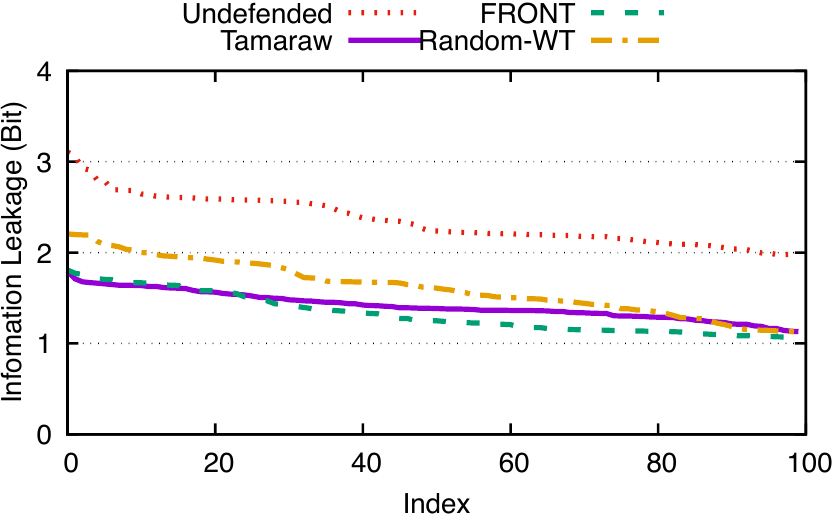}
	\end{center}
	\caption{Top 100 most informative features (indexed by rank) estimated on each dataset.}
	\label{fig:infoleak} 
\end{figure}

Since WF attacks evolve quickly, it is inadequate to demonstrate the effectiveness of a defense merely by showing the low accuracy of existing attacks against it.
Therefore, we adopt the WeFDE technique~\cite{LiGH18} to directly measure the information leakage of defended network traces.  
WeFDE includes 3043 features from website fingerprinting literature and quantifies the information leakage of each feature, defined as the mutual information obtained through a specific feature. 

We perform information leakage analysis on four datasets. 
Since a lot of features share redundant information leakage with the other features, we first remove redundant features and pick the 100 most informative non-redundant features.
Figure~\ref{fig:infoleak} shows the results.
When there is no defense, the information leaked by those features ranges from 1.97 to 3.11 bits. 
For Random-WT, the information leaked ranges from 1.12 to 2.21 bits.
For Tamaraw, the features leak 1.13 to 1.80 bits of information.
FRONT leaks less information (1.06 to 1.51 bits) for the last 76 features, but it leaks more information for the first 24 features (1.55 to 1.81 bits).

\subsection{Open-World Scenario}
\label{sec:open-world-scenario}

Next, we look into the more realistic open-world scenario.
We first investigate an easier binary version of the open-world website fingerprinting problem: 
the attacker is only required to tell whether the client is visiting a page from the monitored list or not, 
without having to determine which specific monitored page it is.
Then we discuss the harder multi-class problem where the attacker should further decide which specific webpage the client is visiting, if it is a monitored one.

\paragraph{Binary-Class Problem}
We use TPR and FPR as the two main metrics for attack evaluation.
As in previous literature~\cite{SirinamIJW18, Cadena20Traffic, RimmerPJGJ18Automated}, we compute the receiver operating characteristic (ROC) curve of each attack for every defense. 
The curve is derived by varying a classifier's confidence threshold of outputting a positive prediction so that the classifier can trade off between TPR and FPR. 
We use the area under the curve (AUC) to evaluate the ability of the classifier to trade off true positives (monitored pages) and negatives (non-monitored pages).
AUC = 1 represents a perfect classifier while AUC = 0.5 represents random guessing. 
For simplicity, we only present the ROC curve of DF (the best attack in our evaluation) against all the defenses here. 
We put the results for the other attacks in Appendix~\ref{app:roc}.

\begin{figure}[]
	\begin{center}
	\includegraphics[width=0.9\linewidth]{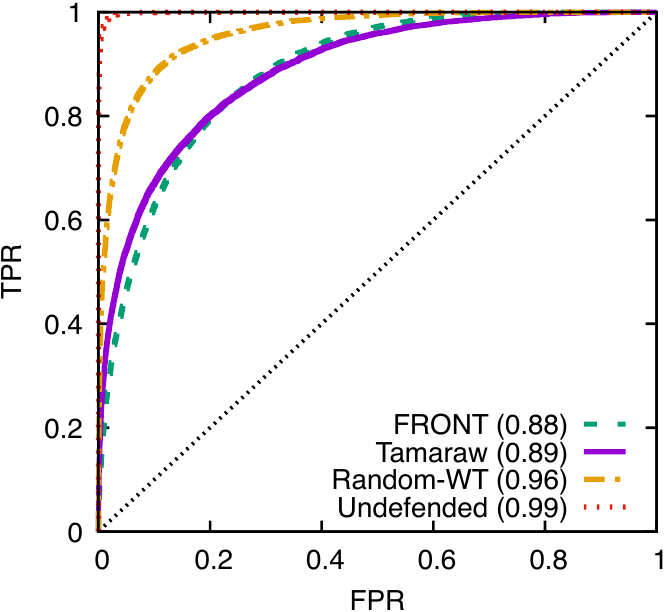}
	\end{center}
	\caption{The ROC curves of DF on different datasets. The diagonal line represents random guessing. The AUC values are given in the legend; a larger value represents better classification performance.}
	\label{fig:roc_of_df} 
\end{figure}

Figure~\ref{fig:roc_of_df} shows that DF achieves 0.99 AUC on the undefended dataset, nearly approaching the perfect classifier.
Its AUC only decreases by 0.03 on the Random-WT dataset.
This means Random-WT fails to make monitored traces indistinguishable from non-monitored ones. 
FRONT and Tamaraw, achieve similar results: the AUC of DF is 0.88 and 0.89, respectively. 
Although they are more effective than Random-WT, DF is still far better than random guessing against these two defenses.

\paragraph{Multi-Class Performance}
\begin{table*}[]
\centering
\caption{Evaluation of the implemented defenses in the open-world scenario of 100,000 traces (multi-class problem). 
We present TPR, FPR, and precision ($\pi$) of each attack.
Tamaraw is the most effective defense against all the attacks, followed by FRONT.
Random-WT is not effective against DF.}
\label{tab:ow-imp-res}
\resizebox{\textwidth}{!}{%
\begin{tabular}{|c||ccc|ccc|ccc|ccc|}
\hline
\multirow{2}{*}{\textbf{Defense $\mathcal{D}$}} &
  \multicolumn{3}{c|}{\textbf{kFP~\cite{Hayes16kfin}}} &
  \multicolumn{3}{c|}{\textbf{CUMUL~\cite{Panchenko16Web}}} &
  \multicolumn{3}{c|}{\textbf{DF~\cite{SirinamIJW18}}} &
  \multicolumn{3}{c|}{\textbf{Var-CNN~\cite{bhat2019var}}} \\
                         & TPR(\%) & FPR(\%) & $\pi(\%)$ & TPR(\%) & FPR(\%) & $\pi(\%)$ & TPR(\%) & FPR(\%) & $\pi(\%)$ & TPR(\%) & FPR(\%) & $\pi(\%)$ \\ \hline \hline
Undefended               & 54.50   & 0.09    & 96.67     & 79.38   & 2.21    & 79.55     & 97.71   & 0.49    & 94.56     & 92.74   & 0.39    & 94.36     \\
Tamaraw~\cite{CaiNWJG14} & 0.96    & 0.02    & 60.38     & 1.21    & 9.47    & 1.18      & 6.43    & 1.79    & 12.02     & 2.44    & 3.76    & 3.75      \\
FRONT~\cite{GongW20}     & 0.87    & 0.01    & 88.78     & 3.55    & 4.14    & 7.46      & 42.79   & 2.49    & 56.58     & 19.02   & 2.45    & 36.57     \\
Random-WT~\cite{WangG17} & 9.63    & 0.06    & 91.19     & 32.55   & 6.25    & 33.70     & 83.41   & 2.81    & 75.13     & 52.01   & 1.69    & 74.16     \\ \hline
\end{tabular}%
}
\end{table*}

Table~\ref{tab:ow-imp-res} presents the full evaluation results for the multi-class problem. 
Besides TPR and FPR, we also present the \textit{precision} (denoted as $\pi$) of each attack.
Precision shows how confident the attacker is about his predictions that classify the visits as monitored. 
When there is no defense, DF is the best attack, achieving 98\% TPR and 95\% precision.
Var-CNN also achieves over 92\% TPR and 94\% precision. 
CUMUL has both 80\% TPR and precision. 
Although kFP has the lowest TPR (55\%), it has the highest precision (97\%). 

Tamaraw lowers the TPR of all attacks to nearly 0\%, except for DF, which has a TPR of 6\%. 
While kFP can still achieve a precision of 60\%, its performance is still poor since it classifies almost every instance to be non-monitored.
The best attack against FRONT is DF, which has a TPR of 43\% and a precision of 57\%. 
Notably, FRONT is rather effective against kFP and CUMUL, approaching the performance of Tamaraw.
kFP achieves less than 1\% TPR against FRONT. 
But compared with Tamaraw, FRONT incurs much less data overhead. 
We find that Random-WT is ineffective against deep learning attacks, since DF can achieve 83\% TPR and 75\% precision against it. 
This shows that, without burst molding, the effectiveness of Walkie-Talkie will be greatly decreased, even though Random-WT seems to incur a higher overhead than Walkie-Talkie.

Overall, we find that the two deep learning attacks outperform kFP and CUMUL with a much higher TPR in all cases. 
This was already known in the undefended and simulated cases; our results confirm that this for real implementations.
DF consistently becomes the best attack against all the defenses,
but kFP is the most precise attack in general.

\paragraph{Balance of Training Set} 
\begin{figure}[]
	\vspace{-37pt}
	\begin{center}
		\includegraphics[width=0.85\linewidth]{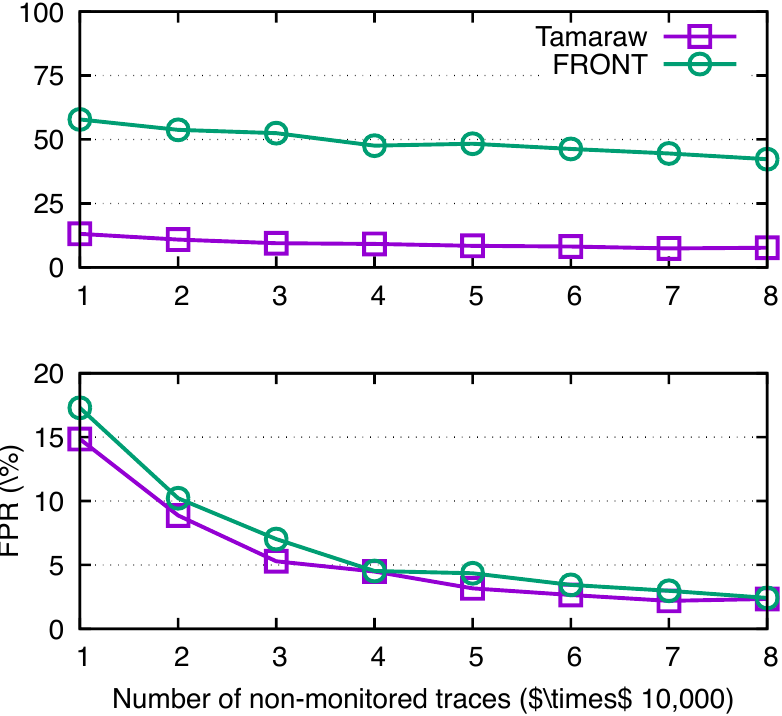}
	\end{center}
	\caption{The performance of DF against two defenses, varying the number of non-monitored traces in the training set. TPR and FPR are both decreasing.}
	\label{fig:ow-vary-unmon} 
\end{figure}
We also want to see how the number of non-monitored training traces could affect classifier performance against WF defenses.
This number is chosen by the attacker as a potential way to improve the attack's performance.
We focus on two state-of-the-art defenses, FRONT and Tamaraw. 
We still perform 10-fold cross validation, but we vary the number of non-monitored instances in the training set from 10,000 to 80,000 while keeping the size of testing set at 10,000 on each fold. 
(In all other experiments, there are 81,000 non-monitored instances in the training set.)
Figure~\ref{fig:ow-vary-unmon} shows the attack performance of DF.

The TPR of DF on the FRONT dataset decrease slowly from 58\% to 42\%.
On the Tamaraw dataset, the TPR remains at a low level, decreasing from 13\% to 8\%.
We notice that there is a dramatic drop of FPR when the number of non-monitored traces in the training set is increased from 10,000 to 40,000. 
In the end, the FPR on both datasets is less than 3\%.
This shows that while the attacker can somewhat increase the TPR by using fewer non-monitored traces in training,
the sharp increase of FPR may not be worth it.

\paragraph{Overhead} 
\begin{table}[]
\centering
\caption{The data and time overhead of each defense in the open-world scenario (100,000 traces).}
\label{tab:defense-imp-open-ovhd}
\resizebox{0.98\linewidth}{!}{%
\begin{tabular}{|c||c|c|}
\hline
\textbf{Defense $\mathcal{D}$} & \textbf{Data Overhead (\%)} & \textbf{Time Overhead(\%)} \\ \hline \hline
Tamaraw~\cite{CaiNWJG14} & 179 & 21 \\
FRONT~\cite{GongW20}     & 75  & 0 \\
Random-WT~\cite{WangG17} & 88  & 23 \\ \hline
\end{tabular}%
}
\end{table}
We recompute the overhead for each defense after we include all traces from non-monitored pages. 
Each dataset is 10 times larger than the closed-world one. 
As shown in Table~\ref{tab:defense-imp-open-ovhd}, we find that all defenses have a slightly higher data overhead and a lower time overhead compared with the closed-world scenario.
Tamaraw is the most expensive defense, incurring 179\% data overhead and 21\% time overhead.  
Random-WT has a similar time overhead as Tamaraw, 
and its data overhead is about half of Tamaraw. 
FRONT is the cheapest defense, with 75\% data overhead and 0\% overhead. 
One possible reason accounting for such a difference is that there are smaller webpages in the non-monitored list,
which only take a few seconds to load whether or not the client uses a defense, leading to a smaller time overhead.
These pages generate few real packets, thus increasing the portion of dummy data and thus the data overhead.

\section{Simulation Evaluation}
\label{sec:simulation-evaluation}
In this section, we simulate each defense on the undefended dataset.
Our goal is to compare implementation and simulation results in the open-world scenario based on overhead and attack performance.
Since DF consistently achieves the best performance, we use DF as the default attack in this section. 

\subsection{Simulating Tamaraw}
\label{sec:simulating-tamaraw}

Previous work had evaluated Tamaraw with simulation only. 
Its protocol is simple: both sides send packets at a constant rate. 
To achieve this, our simulation divides the timeline into equal time slots. 
In each slot, a real or dummy packet is sent, depending on whether or not there is any data in the buffer. 
After the last real packet is sent, the defense will continue to send dummy packets until the length of the packet sequence is some multiple of a parameter $L$. 

To simulate Tamaraw over an undefended dataset, one question is to decide which time slot each real packet should be assigned to. 
Intuitively, each packet should be assigned to a time slot that is \textbf{no earlier} than its original timestamp,
and two packets cannot be assigned to the same time slot.
However, we may need to delay a packet further due to a possible dependency relationship between packets (of both directions);
this dependency relationship is not known in simulation. 
Therefore, we define two simulation strategies here:
\begin{itemize}
	\item \textbf{Optimistic Strategy:} we assume that each packet is informationally independent of each other.
	The packets from each direction will be assigned to the closest time slot that is after their original timestamp. 
	This implies that the packets of different directions may be re-ordered by the defense.
	\item \textbf{Pessimistic Strategy:} we assume that each packet is dependent on all previous packets from the other direction. 
	The packets will then be assigned to the closest time slot that is after their original timestamp while keeping the original packet order. 
\end{itemize}
These two strategies represent two extreme cases for Tamaraw. 
The overhead values we get from the pessimistic strategy would be larger than those from the optimistic strategy. 
Note that these two strategies are not guaranteed to be any upper bound or lower bound of the real situation.
We conduct the simulation with both strategies and give a range of performance results. 
The original work analyzed Tamaraw using the optimistic strategy~\cite{tamarawcode};
to the best of our knowledge, our work is the first to analyze the pessimistic strategy. 

We first compare simulation and implementation under the same parameter settings. 
Table~\ref{tab:cmp-tamaraw} shows the results. 

\begin{table}[]
\centering
\caption{Comparison between simulation and implementation of Tamaraw. We simulate Tamaraw with the optimistic and pessimistic strategies.}
\label{tab:cmp-tamaraw}
\resizebox{\linewidth}{!}{%
\begin{tabular}{@{}ccccc@{}}
\toprule
\multirow{2}{*}{\textbf{}} & \multicolumn{2}{c}{\textbf{Overhead (\%)}} & \multicolumn{2}{c}{\textbf{DF Performance (\%)}} \\ \cmidrule(l){2-5} 
                                                            & Data & Time & TPR  & FPR  \\ \midrule
Optimistic (Section~\ref{sec:simulating-tamaraw})  & 207  & 19   & 2.59 & 1.35 \\
Pessimistic (Section~\ref{sec:simulating-tamaraw}) & 224  & 26   & 2.74 & 1.58 \\
Implementation (Section~\ref{sec:open-world-scenario})      & 179  & 21   & 6.43 & 1.79 \\ \bottomrule
\end{tabular}%
}
\end{table}

\paragraph{Overhead}
 The optimistic simulation has a 207\% data overhead while the pessimistic one has a 224\% data overhead.
 Interestingly, both the optimistic simulation and the pessimistic simulation yield a higher data overhead than implementation (by 28\% and 45\%, respectively).
 While the constant padding of simulated Tamaraw starts from the beginning and will not stop until page load ends,
 implemented Tamaraw can start late and stop early so that the data overhead is reduced:
 \begin{itemize}
 \item When a Tor Browser first launches, there will not be a stable constant padding from both sides during the start-up stage since the two parties have not yet established a connection.
 \item Padding may stop in the middle of a page load due to timeout events caused by network congestion, broken Tor circuits, a slow response from the web server, etc. This is due to the soft stop mechanism we introduced in Tamaraw's state machine (see Section~\ref{sec:implementing-tamaraw}).
      Once stopped, the machine would still restart later if there is more real data. 
 \end{itemize}

\begin{figure}[]
	\vspace{-28pt}
	\begin{center}
		\includegraphics[width=0.9\linewidth]{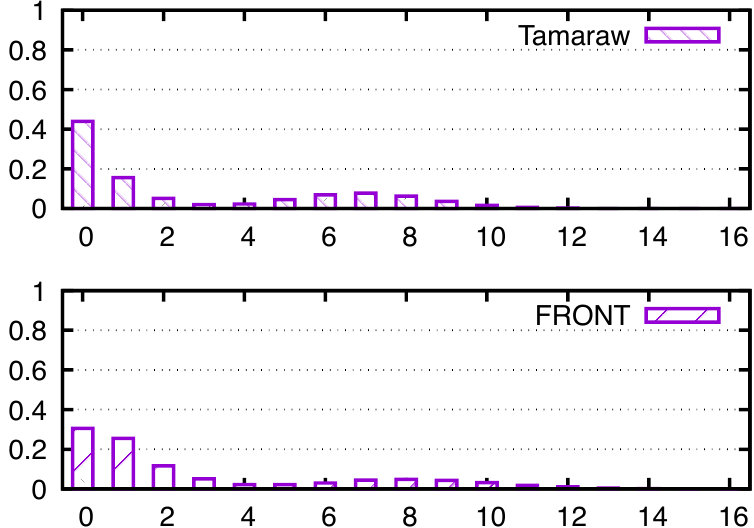}
	\end{center}
	\caption{Normalized frequencies of number of restart events in each loading computed over Tamaraw and FRONT datasets.}
	\label{fig:timeout-analysis} 
\end{figure}

The first case is inevitable for any defense. 
We thus take a deeper look at the second case.
We count the number of restart events in each page load in the dataset in Figure~\ref{fig:timeout-analysis}.
44\% of page loads have no restart events while 16\% have one restart event. 
For the remaining 40\% of page loads where Tamaraw restarts more than once, 96\% come from non-monitored pages,
which are less popular and more likely to load incorrectly. 
We estimate that even if we significantly increase the time window for our soft stop condition from \SI{1}{s} to \SI{5}{s},
38\% of traces would still have at least one restart event. 
However, the data overhead will be greatly increased.

We further investigate the non-monitored traces that have more than 6 restart events and found that most of these contain few real packets.
More specifically, these traces only have 1,327 real packets on average while the number of dummy packets is 9 times more than that of real ones. 
It is likely that these small webpages were gradually loading tiny amounts of data, so they repeatedly triggered our soft stop condition.
In general, most page loads will not frequently trigger stop and restart events under the time window of \SI{1}{s}. 

In terms of time overhead, the result for implementation (21\%) falls in the small range between our simulations (19\% for optimistic, 26\% for pessimistic). 
This is because the default values we set for the parameters are quite small (i.e., a high sending rate) in order to reduce the time overhead of Tamaraw.
 
\paragraph{Attack Performance}
DF achieves similar results for the two simulation strategies: less than 3\% TPR and 2\% FPR. 
By contrast, DF achieves about 6\% TPR and 2\% FPR in implementation. 
It is possible that the aforementioned gaps in the defense leak a small amount of data that is leveraged by DF to increase its TPR slightly;
nevertheless, the best attack in the field still cannot defeat our real implementation of Tamaraw.

\subsection{Simulating FRONT}
\label{sec:simulating-front}
Since FRONT injects dummy data independently of the page loading process, its simulation is simple;
however, differences between the implementation and the simulation can still occur due to network congestion and delays.
We directly use the code provided by the authors of FRONT for our simulation~\cite{wfcode}. 
We compare implementation results with two simulation cases:
\begin{itemize}
	\item \texttt{Sim-1} use the same parameters as implementation where $N_{out} = N_{in} = 3000$, $W_{min} = \SI{1}{s}$ and $W_{max} = \SI{13}{s}$.
	\item \texttt{Sim-2} has the same data overhead as implementation where $N_{out} = N_{in} = 4400$, $W_{min} = \SI{1}{s}$ and $W_{max} = \SI{13}{s}$.
\end{itemize}
Table~\ref{tab:cmp-front} shows the detailed results. 
\begin{table}[]
\centering
\caption{Comparison between simulation and implementation of FRONT. \texttt{Sim-1} uses the same parameters as implementation while \texttt{Sim-2} achieves the same data overhead as implementation.}
\label{tab:cmp-front}
\resizebox{\linewidth}{!}{%
\begin{tabular}{@{}ccccc@{}}
\toprule
\multirow{2}{*}{\textbf{}}                             & \multicolumn{2}{c}{\textbf{Overhead (\%)}} & \multicolumn{2}{c}{\textbf{DF Performance (\%)}} \\ \cmidrule(l){2-5} 
                                                       & Data                 & Time                & TPR                     & FPR                    \\ \midrule
\texttt{Sim-1} (Section~\ref{sec:simulating-front})    & 52                   & 0                   & 72.06                   & 2.16                   \\
\texttt{Sim-2} (Section~\ref{sec:simulating-front})    & 75                   & 0                   & 55.44                   & 2.12                   \\
Implementation (Section~\ref{sec:open-world-scenario}) & 75                   & 0                   & 42.79                   & 2.49                   \\ \bottomrule
\end{tabular}%
}
\end{table}

\paragraph{Overhead}
With the same parameters, the FRONT implementation incurs 23\% more data overhead than \texttt{Sim-1}. 
We examine the log of WFDefProxy and find that FRONT may restart several times during a page load.
The distribution of the number of restart events is similar as that of Tamaraw (see Figure~\ref{fig:timeout-analysis}).
More specifically, 44\% of page loads have more than one restart event:
among them, over 90\% come from non-monitored pages. 
Unlike Tamaraw, this leads to more padding than expected, as FRONT adds more padding at the front of each page load. 

The additional triggers may be due to the latency of the network or occasional timeouts during page loading, as discussed in Section~\ref{sec:simulating-tamaraw}.
We also find that 45\% of all restart events happen after more than 80\% of the real packets are delivered (around the end of a trace).
That also explains why the real data overhead (75\%) is close to the simulated value (52\%), 
as most of the additional packets triggered by a restart will be discarded at the end. 

\paragraph{Attack Performance}
We first compare \texttt{Sim-1} with implementation: 
we find that DF achieves over 72\% TPR against \texttt{Sim-1}, 29\% higher than implementation, with similar FPR (2\%).
One reason is that the implementation injects 23\% more dummy data than simulation,
but this is not the only reason, as the following experiment shows.
We next compare \texttt{Sim-2} with implementation:
we find that the implementation of FRONT is still more effective than its simulation even with the same data overhead. 
DF achieves 43\% TPR and 2\% FPR in reality while it has 55\% TPR and 2\% FPR in simulation. 
This indicates that FRONT can benefit from the random nature of the network;
it becomes even more effective when the noise can be perturbed by long periods of downtime.

\subsection{Simulating Random-WT}
\label{sec:simulating-random-wt}
Random-WT pads real bursts and injects dummy bursts with probability $p_{fake}$.
To simulate Random-WT in a more accurate way, we first collect a new undefended dataset where the websites are loaded in half-duplex mode at the network layer (introduced in Section~\ref{sec:implementing-random-wt}).
Then, we pad the real bursts and inject fake bursts according to the protocol. 
The biggest challenge is how we simulate the delay introduced by the defense. 
Here we make the following assumptions:
\begin{itemize}
	\item When we pad a real burst, we assume the dummy packets padded in the end are sent out together with the real packets without introducing any delay.
	\item To simulate the round trip time (RTT) between a fake request burst and a fake response burst, we measure the latest RTT $\tau$ between real bursts and then sample a fake RTT $\tau_{fake}$ from a normal distribution with mean $\tau$ and variance $0.1\tau$.
\end{itemize}

To ensure that the fake bursts look similar to the padded real bursts, we must pad less data on the real bursts than the fake ones.
As described in Section~\ref{sec:implementing-random-wt}, we mainly need to decide two sets of parameters for Random-WT:
the maximum padding number for real bursts, $N^{real}_{out}$ and $N^{real}_{in}$,
and the maximum padding number for fake bursts, $N^{fake}_{out}$ and $N^{fake}_{in}$.
We first compute the mean size of bursts based on the monitored traces: this was 4 for the outgoing bursts and 45 for the incoming bursts. 
We set $N^{real}_{out} = 4, N^{real}_{in} = 45$, which are exactly the mean values of the bursts on both sides.
We set $N^{fake}_{out} = 8, N^{fake}_{in} = 90$, which are twice that of the mean values,
so that fake bursts have the same mean size as real bursts plus expected padding. 
Then we vary $p_{fake}$ from 0 to 1 and test Random-WT against four attacks in the closed-world scenario. 
When we increase $p_{fake}$, both the time overhead and data overhead increase and the accuracy of the attacks decrease.
The full results are shown in Appendix~\ref{sec:parameter-tuning-for-random-wt}.
We finally choose $p_{fake} = 0.4$ as a trade-off between overhead and attack accuracy. 

\begin{table}[]
\centering
\caption{Comparison between simulation and implementation of Random-WT.}
\label{tab:cmp-random-wt}
\resizebox{\linewidth}{!}{%
\begin{tabular}{@{}ccccc@{}}
\toprule
\multirow{2}{*}{\textbf{}}                             & \multicolumn{2}{c}{\textbf{Overhead (\%)}} & \multicolumn{2}{c}{\textbf{DF Performance (\%)}} \\ \cmidrule(l){2-5} 
                                                       & Data                 & Time                & TPR                     & FPR                   \\ \midrule
Simulation (Section~\ref{sec:simulating-random-wt})    & 96                   & 44                  & 87.47                   & 2.50                  \\
Implementation (Section~\ref{sec:open-world-scenario}) & 88                   & 23                  & 83.41                   & 2.81                  \\ \bottomrule
\end{tabular}%
}
\end{table}
Next, we compare the simulation and implementation results in Table~\ref{tab:cmp-random-wt}.

\paragraph{Overhead}
The data overhead measured over the simulated dataset is 96\%, only 8\% higher than the implementation result. 
This shows that our simulation accurately captures the real data overhead of Random-WT.
However, our simulation overestimates the time overhead by 21\%. 

\paragraph{Attack Performance}
The TPR of DF is similar in simulation (87\%) and in implementation (83\%). 
The FPR difference is only 0.3\%. 
This suggests that our simulation over the undefended dataset (collected in half-duplex mode) is accurate.

\section{Discussion}
\label{sec:discussion}
\subsubsection*{Do WF defenses work in reality?}
Our results show that implemented Tamaraw is still effective against the state-of-the-art attacks,
but its data overhead is quite high. 
FRONT also provides a certain level of security against the best attacks, lowering the TPR of DF by 55\% and Var-CNN by 74\% in the open-world scenario.
Given that FRONT has nearly zero time overhead and is relatively simple to implement, it can be considered as a candidate defense for Tor.
On the other hand, Random-WT is totally ineffective against deep learning attacks. 
Despite a higher cost, Random-WT only reduces the accuracy of DF by 7\% in the closed-world scenario.
Even in the open-world scenario, DF still has 83\% TPR and 75\% precision. 
The failure of Random-WT shows that the half-duplex mode, by itself,
does not provide much security, even with dummy packets to cover burst sizes. 

\subsubsection*{Is simulation accurate enough?}
Simulation allows the researcher to preview how traces will look like under a defense.
It is much easier and quicker to obtain results from simulation than from implementation. 
Simulation does offer some accurate results:
simulated Tamaraw and Random-WT yield the expected TPR and FPR of DF in the open-world scenario. 
However, simulation forces us to make assumptions about page loading, such as the client/proxy knowing when page load starts and ends.
It is interesting to find that relaxing these assumptions in our implementation does not actually harm the performance of these defenses.
As a lightweight defense focusing on leveraging randomness,
FRONT even achieves 13\% less TPR in implementation than simulation with the same overhead.

On the other hand, it is hard to accurately predict the overhead through simulation. 
For example, FRONT incurs 23\% more data overhead in implementation than simulation because FRONT padding can be unexpectedly triggered more than once in each page load. 
Another reason is that the random nature of the real network brings about more uncertainty. 
It is challenging to simulate network latency without knowing network conditions and the dependency relationship between packets.
Regularized defenses can trigger timeouts in other network layers, leading to retransmission and changes in traffic patterns. 
We find that even at a high packet rate, the simulation of Tamaraw overestimates data overhead by 28\% -- 45\%.
The time overhead of Random-WT is overestimated by 21\%.

While the use of simulation is valuable for preliminary results and for exploring parameters quickly,
our work shows that their overhead and attack performance results can both be inaccurate for unpredictable reasons,
and we should ultimately analyze defenses through implementation. 

\section{Conclusion and Future Work}
\label{sec:conclusion}
In this paper, we proposed and implemented WFDefProxy, a pluggable transport that can serve as a platform for empirically evaluating WF defenses on Tor.
We are the first to fully deploy Tamaraw, FRONT and Random-WT, overcoming previously made assumptions in the simulation environment. 
All of our defenses are usable by any Tor users who wish to set up a bridge or connect to an existing bridge running our code.
For each defense, we crawled a dataset containing 100,000 traces and conducted an extensive evaluation in both closed-world and open-world scenarios.
Results show that Tamaraw can effectively lower the TPR of all attacks to less than 7\% in the open-world scenario. 
We found that FRONT is much more effective than Random-WT. 

To compare simulation and implementation results, we simulated each defense over the undefended dataset and compared the results with our implementation.
Our implementation of FRONT incurred 23\% more data overhead than simulation
while our implementation of Tamaraw incurred 28\% -- 45\% less data overhead (depending on the simulation strategy used). 
We encourage others to use this platform to evaluate existing WF defenses as well as future ones. 

To be able to implement these defenses on Tor in a modular and easily installed manner,
we had to implement WFDefProxy as a pluggable transport, which implies that the bridge is a trustable party in our threat model.
In reality, the bridge could be a potential WF attacker. 
Therefore, it would be better to deploy the defenses on the middle node in practice.
We leave this as future work pending changes to the Tor protocol that allow pluggable transports to be installed on the middle node.

%

\section*{Availability}
We publish all our codes in this paper: 
\begin{itemize}
	\item Codes for WFDefProxy:
	
	\url{https://anonymous.4open.science/r/wfdef-11EF/}
	\item Codes for WFCrawler, a toolkit working together with WFDefProxy for crawling and parsing traces:
	
	\url{https://anonymous.4open.science/r/WFCrawler-D46B/}
	\item Simulation codes for Tamaraw and Random-WT:
	
	\url{https://anonymous.4open.science/r/SimWFDef-8D7C/}
	
\end{itemize}

We will also upload all of our collected datasets upon publication. 

\bibliographystyle{plain}
\bibliography{main.bib}

\begin{appendices}
\section{ROC Curve for Each Attack}
\label{app:roc}
\begin{figure*}[]
   \centering
   \begin{subfigure}{.325\textwidth}
     \centering
     \includegraphics[width=0.9\linewidth]{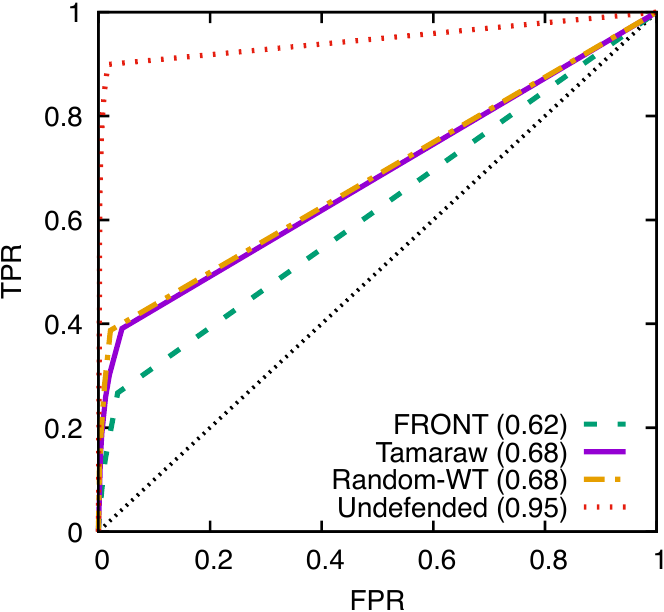}
     \caption{kFP}
   \end{subfigure}
   \hspace{1pt}
   \begin{subfigure}{.325\textwidth}
     \centering
     \includegraphics[width=0.9\linewidth]{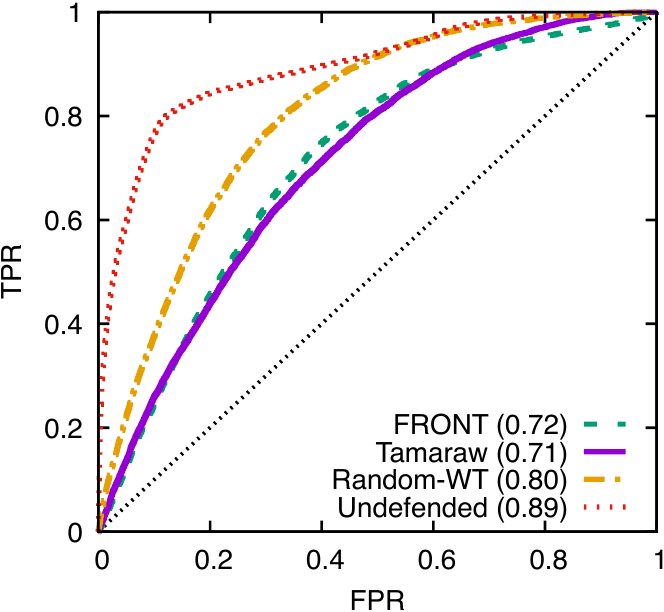}
     \caption{CUMUL}
   \end{subfigure}
    \hspace{1pt}
   \begin{subfigure}{.325\textwidth}
     \centering
     \includegraphics[width=0.9\linewidth]{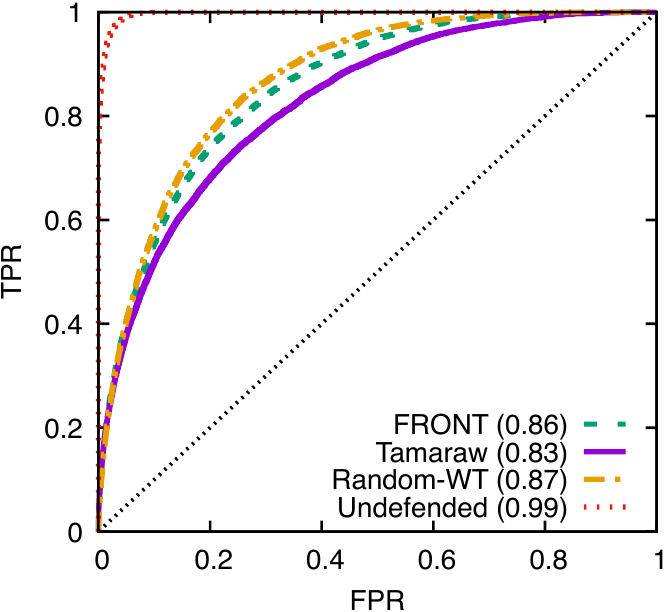}
     \caption{Var-CNN}
   \end{subfigure}
   \caption{The ROC curves of three attacks against different defenses. The diagonal line represents random guessing. The AUC values are given in the legend; a larger value represents better classification performance.}
   \label{fig:roc-curve-app}
\end{figure*}

We already present the ROC Curve of DF for binary problem in the open-world scenario in~\ref{sec:open-world-scenario}.
In the section, we present the results for other three WF attacks.
For kFP, the way to derive the ROC curve is slightly different from the other attacks. 
We follow the way introduced in~\cite{SirinamIJW18}: 
we find $k=6$ closest neighbours and output the prediction probability as the fraction of neighbours that have the same label as the test instance. 

As is shown in~\ref{fig:roc-curve-app}, Var-CNN achieves the best performance among the attacks on every dataset. 
When there is no defense, both Var-CNN and kFP achieve an AUC of more than 0.95 while CUMUL only achieves 0.89 AUC. 
However, CUMUL has higher AUC on the other datasets compared with kFP.
kFP is especially vulnerable to FRONT with an AUC of only 0.62. 
This is because kFP relies heavily on FRONT features. 

\section{Parameter Tuning for Random-WT}
\label{sec:parameter-tuning-for-random-wt}
\begin{table}[]
\centering
\caption{The performance of Random-WT with different $p_fake$. $N^{real}_{out}, N^{real}_{in}, N^{fake}_{out}$ and $N^{fake}_{in}$ are fixed at 4, 45, 8 and 90, respectively. We mark the parameters used in implementation in bold font.}
\label{tab:random-wt-vary-p}
\resizebox{\linewidth}{!}{%
\begin{tabular}{|c|cc|c|c|c|c|}
\hline
\multirow{2}{*}{\textbf{$p_{fake}$}} & \multicolumn{2}{c|}{\textbf{Overhead (\%)}} & \multicolumn{4}{c|}{\textbf{Attack Accuracy(\%)}}                                                         \\ \cline{2-7} 
                                     & Data                 & Time                 & kFP~\cite{Hayes16kfin} & CUMUL~\cite{Panchenko16Web} & DF~\cite{SirinamIJW18} & Var-CNN~\cite{bhat2019var} \\ \hline
0                                    & 48                   & 24                   & 76.65                  & 81.58                       & 96.31                  & 92.31                     \\
0.2                                  & 68                   & 42                   & 70.29                  & 74.26                       & 94.89                  & 86.8                      \\
\textbf{0.4}                         & \textbf{88}          & \textbf{59}          & \textbf{68.56}         & \textbf{69.89}              & \textbf{93.65}         & \textbf{86.45}            \\
0.6                                  & 108                  & 76                   & 67.12                  & 67.01                       & 93.16                  & 82.25                     \\
0.8                                  & 128                  & 94                   & 66.38                  & 64.95                       & 92.58                  & 82.9                      \\
1                                    & 148                  & 111                  & 65.81                  & 64.74                       & 91.67                  & 82.23                     \\ \hline
\end{tabular}%
}
\end{table}
Since Random-WT has five parameters and there is no suggested values for them, we need to first tune the parameters to see how these parameters affect the performance of the defense. 
Given that the search space for the parameters is very large, 
we first set $N^{real}_{out} = 4, N^{real}_{in} = 45, N^{fake}_{out} = 8$ and $N^{fake}_{in} = 90$ and vary $p_{fake}$ from 0 to 1. 
The idea here is to make the size of fake bursts similar to that of real burst after the padding. 
With such $N^{real}_{out}$ and $N^{real}_{in}$, we expect to have about 50\% dummy packets added into the real bursts. 
Table~\ref{tab:random-wt-vary-p} shows the accuracy of each attack against Random-WT with every setting.

Both the data overhead and the time overhead increase when $p_{fake}$ is increased because there is a higher chance to add a dummy burst. 
We also find that when $p_{fake}$ changes from 0 (no fake bursts) to 0.2 (may have fake bursts), there is a relative big drop in accuracy for all the attacks, indicating that it is necessary to insert fake bursts for Random-WT. 
But when $p_{fake} > 0.6$, the accuracy of each attack barely decreases while the overheads still grow linearly. 
Therefore, we choose $p_{fake} = 0.4$ as our default setting in our implementation. 
\end{appendices}

\end{document}